\documentclass[sigconf]{acmart}

\setcopyright{acmlicensed}
\copyrightyear{2025}
\acmYear{2025}
\acmDOI{XXXXXXX.XXXXXXX}

\usepackage{graphicx,url}
\usepackage[utf8]{inputenc}  
\usepackage{color, colortbl}

\usepackage{hyperref}

\begin{document}
%
\title{M\&SCheck: Towards a Checklist to Support Software Engineering Newcomers to the Modeling and Simulation Area}

\author{Luiza Martins de Freitas Cintra}
\email{cintraluiza@discente.ufg.br}
\affiliation{%
  \institution{Federal University of Goiás}
  \city{Goiânia}
  \country{Brazil}
}

\author{Philipp Zech}
\email{philipp.zech@uibk.ac.at}
\affiliation{%
  \institution{University of Innsbruck}
  \city{Innsbruck}
  \country{Austria}
}

\author{Mohamad Kassab}
\email{mkassab@bu.edu}
\affiliation{%
  \institution{University of Boston}
  \city{Boston}
  \country{USA}
}

\author{Eliomar Araújo Lima}
\email{eliomar.lima@ufg.br}
\affiliation{%
  \institution{Federal University of Goiás}
  \city{Goiânia}
  \country{Brazil}
}

\author{Sofia Larissa da Costa Paiva}
\email{sofialarissa@ufg.br}
\affiliation{%
  \institution{Federal University of Goiás}
  \city{Goiânia}
  \country{Brazil}
}

\author{Valdemar Vicente Graciano Neto}
\email{valdemarneto@ufg.br}
\affiliation{%
  \institution{Federal University of Goiás}
  \city{Goiânia}
  \country{Brazil}
}


\begin{abstract}
The advent of increasingly complex and dynamic ecosystems, such as digital twins (DT), smart cities and Industry 4.0 and 5.0, has made evident the need to include modeling and simulation (M\&S) in the software development life cycle. Such disruptive systems include simulation models in their own architecture (such as DT) or require the use of simulation models to represent the high degree of movement and the multiplicity of interactions that occur between the involved systems. However, when software engineers (particularly the newcomers) need to use M\&S in their projects, they often pose themselves an important question: which formalism should I use? In this direction, the main contribution of this paper is the establishment of a preliminary checklist with questions to assist beginners in M\&S in choosing the most appropriate paradigm to solve their problems. The checklist is based on three main formalisms: DEVS, System Dynamics and Agent-Based Simulation. A pilot study was carried out and an expert was consulted. The preliminary results show (i) conformance between the suggestion given by the checklist and the formalism selected in the original studies used as input for evaluating the checklist, and (ii) a positive feedback from the expert.
\end{abstract}

\ccsdesc[500]{Software and its engineering ~ Software development methods}
\ccsdesc[300]{Software and its engineering ~ Software creation and management}
\ccsdesc{Computing methodologies ~ Modeling and simulation}
\ccsdesc[100]{Computing methodologies ~ Modeling methodologies}

\keywords{Modeling, Simulation, DEVS, Agent-Based Simulation, Systems Dynamics}

\maketitle

\section{Introduction}

Modeling and Simulation (M\&S) is an area dedicated to the use of simulation models, usually supported by animation, to exercise scenarios and predict properties of systems of interest \cite{singh2009system}. In Software Engineering (SE), it is possible to find several applications of M\&S in: (i) \textit{critical domains}, in which it is unfeasible to use a real system to predict adverse conditions or specific properties—either because it is too costly or because errors may cause damage and losses \cite{schlager2008hardware}; (ii) \textit{predicting structure and behavior still at design-time} \cite{singh2009system}; or (iii) \textit{planning, management, and decision-making support}, by simulating products or processes, estimating risks and other metrics, and making decisions based on predictions \cite{de2021opportunities}. We can also witness simulation in the software development life cycles, particularly when engineering a Digital Twin (DT), in which the virtual replica can be itself a simulation model acommodated in the DT architecture \cite{agalianos2020discrete,liu2024ai,santos2024,DEPAULAFERREIRA2020}.

A recent study conducted with software engineers showed that simulation formalisms are not aligned with the daily practice of these professionals \cite{Lebtag2022}. Among 58 participants, more than a half (30 participants) had no prior contact with M\&S, although several of them recognized advantages and opportunities for its application in SE, such as architecture evaluation and documentation. Moreover, an unexperienced SE professional would need a considerable time to read papers and execute tutorials towards getting familiar with those formalisms in order to eventually choose one of them. In this context, when those professionals wish to join the M\&S field, one of the first questions that arise from them is: \textit{Which formalism should I use given the characteristics of the problem I have to solve?}

Although this question can sound as trivial for experienced M\&S researchers and practitioners (R\&P), for newcomers this can be concerning. Newcomers face a world with many simulation formalisms and their variants, besides dozens of tools that implement those formalisms. Moreover, the newcomer needs to recognize the characteristics of the problem s/he has in hand to solve and try to match those characteristics to the support offered by the available formalisms, which s/he often is not aware. 

Given those difficulties, the main contribution of this paper is to provide a preliminary checklist to assist beginners in M\&S in choosing a paradigm that can fit the characteristics of the problem they intend to solve, based on three well-known formalisms in the field \cite{Franca2013}: DEVS, System Dynamics, and Agent-Based Simulation. A pilot study was conducted based on scenarios, and an expert was consulted about the checklist. Preliminary results suggest that, when using the checklist to assess scenarios described in previous studies, the choice indicated by the checklist was the same as that adopted in each evaluated study.

This paper is organized as follows: Section 2 presents a short background and related work; Section 3 describes the research method; Section 4 presents the checklist; Section 5 reports the evaluation and results; and Section 6 concludes the paper.

\section{Background}
M\&S enables the study of complex systems without intervening on them directly, investigating properties, impacts, anticipation of losses, and detection of anomalies. Common paradigms include (i) Discrete-Event Simulation (DES) \cite{Zeigler72}, (ii) System Dynamics \cite{forrester1961industrial}, and (iii) Agent-Based Simulation (ABS) \cite{Franca2013}, identified by França and Travassos (2013) as among the most used in SE.

\textbf{Discrete-Event Simulation} (based on the \textit{Discrete-Event Systems Specification}, DEVS) supports M\&S of systems with state changes triggered by discrete events. Created in the 1970s, DES has a solid formal foundation \cite{Zeigler72,Zeigler2016BookGuideMSSoS} and is hierarchically structured into \textit{atomic models} (indivisible components) and \textit{coupled models} (compositions of atomic and coupled models), following the \texttt{Composite} design pattern. \textbf{System Dynamics}, emerged in the mid-20th century alongside with General Systems Theory, is characterized by \cite{Kirkwood:1994:CPM}:
(i) \textit{Feedback Loops}, where variables influence each other cyclically; (ii) \textit{Stocks and Flows}, with stocks as quantifiable resources (e.g., employees, students) and flows as variables that regulate their variation; and (iii) \textit{Goals}, representing target levels the system seeks to reach. Finally, \textbf{Agent-Based Simulation} (ABS) emphasizes the explicit representation of entities and their interactions with each other and the environment. In ABS, \textit{agents} are autonomous entities with decision-making capacity and random responses, such as bacteria, viruses, humans, animals, plants, cars, and organizations. 

\noindent\textbf{Related Work.} Checklists have long been used in Software Engineering to improve quality, transparency, and reproducibility in both research and practice. A notable example is the framework by Petersen et al. (2021) \cite{PETERSEN2021103541}, which systematizes the reporting of contextual factors to enable replication of software development scenarios made in industry. In contrast, our work targets a complementary gap: helping software engineering newcomers select among simulation paradigms based on problem characteristics. Another line of research focuses on how M\&S can be more effectively adopted and used in both academia and industry. França and Ali \cite{de2020role} focus on software engineering and argue that simulation, despite its recognized potential, has not been widely adopted due to unclear expectations, lack of methodological support, and dispersed knowledge on model building and calibration.  Their contribution lies not only in advocating for the use of M\&S but in providing structured methodological support to facilitate its adoption by the software engineering community. Together, these works converge on the idea that supporting adoption of M\&S involves methodological structuring and usability enhancements, offering practical pathways to bridge the gap between potential and effective use of simulation in both research and industry. The next section discusses the research method used to establish the proposed checklist.

\section{Research Method}

For conceiving the checklist, a research method was structured into well-defined steps, as follows. Firstly, we carried out a (i) \textbf{Exploratory Study}, in which an exploratory literature review was conducted to refine the knowledge on the three main formalisms; after, a step on (ii) \textbf{Appropriation and Practice with Tools of each formalism} (DEVS, Agents, and System Dynamics) took place, in which the authors executed tutorials and developed \textit{toy examples} to become familiar with each formalism and understand their particularities, characteristics, and common fields of application. For DEVS, the MS4Me software was used\footnote{\url{https://ms4systems.com/home}}; for Agents, NetLogo\footnote{\url{https://ccl.northwestern.edu/netlogo/}}; and for System Dynamics, VenSim\footnote{\url{https://vensim.com}}. Later, (iii) a \textbf{brainstorming session inspired by focus group practices} was conducted, in which the authors identified the main characteristics of the problems typically solved with each of the studied formalisms. Subsequently, the (iv) \textbf{elaboration of the checklist} was performed, structured into a set of nine criteria to be considered when recommending one of the three formalisms for a project that requires the use of M\&S. After that, a (v) \textbf{pilot study} was carried out; in that study, the checklist was practically applied in scenarios commonly described in the specialized literature. After, the (vi) \textbf{expert evaluation} happened; and, finally, the (vii) \textbf{communication of results}. Expert evaluation will progress in the near future, and this paper is part of step (vii, communication of results), which will also have further developments, as this is an ongoing study. The next section describes the product of step (iv, elaboration of the checklist), as well as the conduction (v, pilot study) and preliminary results of step (vi). It is important to remark that two of the coauthors have more than 10 years of experience with M\&S. 

\begin{table*}[!ht]
\centering
\small
\caption{Simulation Checklist.}

\begin{tabular}{lp{10.0cm}lp{10.7cm}|}

\hline {\textbf{Adherence to the Problem you have in hand} } & {\textbf{Characteristics} } \\

\hline 
(  ) Low
(  ) Medium 
(   ) High
& C1. \textbf{Fidelity.} The simulated counterpart of the real-world entities demand high fidelity, with fine granularity and details in the abstraction.  \\

(  ) Low
(  ) Medium 
(   ) High & C2. \textbf{Human Component.} The system being represented includes a human component, which increases the degree of uncertainty and randomness.\\

(  ) Low 
(  ) Medium 
(   ) High & C3. \textbf{Dynamic Interactions.} The interactions represented in the simulation cause substantial changes in the entities, and the entities’ behavior affects the environment being represented. \\

(  ) Low 
(  ) Medium 
(   ) High & C4. \textbf{Discrete Behavior.} The decisions and behaviors of the entities can be discretized, i.e., they are non-continuous. \\

(  ) Low
(  ) Medium
(   ) High & C5. \textbf{Atomic Structure.} The structure of the entities being represented is indivisible (atomic).\\

(  ) Low
(  ) Medium
(   ) High & C6. \textbf{Composite Structure.} The structure of the entities being represented can be composed of other parts, including structural changes over time. \\

(  ) Low
(  ) Medium
(   ) High & C7. \textbf{Heterogeneity.} The real-world entities being represented in the simulation are heterogeneous. \\

(  ) Low
(  ) Medium
(   ) High & C8. \textbf{Event Sequence.} The data primarily involve a sequence of events.\\

(  ) Low
(  ) Medium
(   ) High & C9. \textbf{Feedback.} The elements of the system under study have causal relationships that intrinsically lead to system feedback.\\

\hline
\end{tabular}
\label{checklist}
\end{table*}

\section{A Checklist for Recommending a Simulation Paradigm Based on Problem Characteristics}

The purpose of the checklist is to assist R\&P who need and/or intend to use M\&S in their projects. The checklist can support him/her to choose the paradigm that eventually matches the characteristics of the problem s/he intends to solve. 

\noindent\textbf{Structure and How Use the Checklist.} Table \ref{checklist} presents the conceived checklist. This artifact is structured into nine different problem characteristics. The checklist reflects typical characteristics of problems solved by software-intensive systems with greater or lesser adherence to the M\&S paradigms mentioned earlier (Agents, Discrete Events, and Systems Dynamics). The checklist should be used, as follows:
\\
\noindent \textbf{(1)} The user is presented to the checklist by an applicant. The user considers its own problem at hand to be solved as the basis to use the checklist, \textit{Example: evaluating the impact of urban policies in a smart city scenario} \cite{caragliu2019smart}.
\\
\noindent \textbf{(2)} The user analyzes problems characteristics C1 to C9 and assesses whether each characteristic has low, medium, or high adherence to the problem s/he has in hand to solve.
\\
\noindent \textbf{(3)} After selecting the levels of importance for each of the problems charateristics, the applicant then applies the rules to recommend, at the end, which would be a simulation formalism that fits his/her problem. In case of absense of an applicant, the rules can be made public and the user himself/herself can analyze the intersection between the analyzed problem and the formalism whose characteristics are more adherent after the selection.

 The checklist was designed so that the user selects a set of five characteristics considered to be most adherent to the problem they are going to solve. The expectation is that there will be an intersection between the selected set and the subsets that characterize each paradigm, which would allow recommending one of them to the checklist user as the outcome.  

\textbf{The paradigms and their respective characteristics are: Agents (C1, C3, C4, C7), DES (C2, C4, C5, C6, C8), and System Dynamics (C2, C5, C6, C9)}. In situations where the results are inconclusive—what we refer to as a ``limbo'', the following tie-breaking or weighting factors were defined: (i) Should the simulation reproduce reality with more graphical details/appeal? (ii) Which tool does offer greater ease of use? To address this, a succinct analysis must be conducted to determine which selected points from the checklist carry greater weight and relevance for the problem in question, as well as to consider the available tools and choose the one with the highest usability affinity. The next section presents the results of the pilot study.

\section{Pilot Study}
\label{cenarios}

To evaluate the conceived artifact, two approaches were adopted. Firstly, using simulation scenarios described in the literature and showing how the checklist would help suggest the formalism adopted in the original study. Four scenarios are presented and discussed. Secondly, an expert was consulted. 
\\\\
\noindent\textbf{5.1. Scenario-Based Evaluation.} Scenario-based evaluation has already been employed in other studies in literature \cite{molina:2023,PETERSEN2021103541}. In this first assessment, typical scenarios found in prior work were considered. The premise is that, starting from the scenario described in the previous work, it would be possible to arrive at the same formalism suggested for that context (assuming the selected paradigm for that scenario was the most suitable), as follows. 
\\\\
\noindent\textbf{Scenario \#1}: Smart Cities' Simulation \cite{9245311}.
\\
\noindent\textbf{Brief Description:} An urban manager needs to evaluate the impacts of implementing smart city technologies — such as mobility systems and energy efficiency — to support decisions on urban planning and services.
\\
\noindent\textbf{Formalism used in the original work:} DES.
\\
\noindent\textbf{Checklist Use:} The evaluation shows that a high-fidelity (C1) is not required, since entities like mobility and energy systems are represented at an aggregate level rather than in fine-grained detail. Human component (C2) is excluded because, in this case, people are not modeled. In contrast, interdependence (C3) is central, as the simulation seeks to capture how subsystems such as land use, transport, and communication affect each other. Discrete behavior (C4) can be also relevant because decisions and processes can be represented as finite events. Some systems can be represented as atomic (C5), but composite structures are also needed (C6), since the city is structured composing systems in many levels. Heterogeneity (C7) is evident in the diversity of infrastructures and platforms. Event sequences (C8) are critical because outcomes depend on the progression and interplay of discrete events. Feedback (C9), however, does not apply, as no recursive adaptation is explicitly modeled. Taken together, the strongest criteria are C3 through C8 which align closely with the characteristics of DES and support its suitability for this urban context.

\noindent\textbf{Result:} DES (C3, C4, C5, C6, C7, and C8 High).
\\\\
\noindent\textbf{Scenario \#2}: System dynamics modelling in supply chain management, focusing on inventory policies, demand amplification, and integration strategies \cite{899737}.
\\
\noindent\textbf{Brief Description:} A supply chain manager needs to evaluate the impacts of inventory policies, demand amplification, and supply chain integration strategies on performance and resilience.
\\
\noindent\textbf{Formalism used in the original work:} Systems Dynamics
\\
\noindent\textbf{Checklist Use:} In the supply chain management scenario, fidelity (C1) is of limited importance, since the focus lies on aggregate flows rather than detailed actors. The human component (C2) appears in managerial and policy decisions but only in aggregated form. Interactions (C3) are central, as oscillations in stocks, demand amplification, and systemic dynamics emerge from the interdependence among entities. Discrete behavior (C4) is less prominent, because flows of orders, production, and shipping are better represented as continuous processes. Atomicity (C5) is relevant, with suppliers, distributors, and retailers treated as indivisible units, though compositional change (C6) also matters, as firms restructure, outsource, or adapt processes over time. Heterogeneity (C7) is critical, reflecting the distinct roles and constraints of different supply chain actors, while event sequences (C8) are less dominant since dynamics arise mainly from continuous flows. Feedback (C9) is fundamental, with phenomena such as the bullwhip effect resulting directly from reinforcing and balancing loops between demand, delays, and replenishment. Taken together, the predominance of interactions, atomic and composite structures, heterogeneity, and feedback confirms the suitability of system dynamics for capturing the nonlinear and emergent behavior of supply chains.
\\
\noindent\textbf{Result:} Systems Dynamics (C3, C5, C6, C7, C9 High).
\\\\
\noindent\textbf{Scenario \#3}: Urban mobility and emissions in Singapore
\cite{werlang2013simulaccao}.
\\
\noindent\textbf{Brief Description:} A transportation planner wants to evaluate how land use, mobility policies, and new technologies will impact urban transportation systems and emissions in Singapore over the coming decades \cite{adnan2016simmobility}.
\\
\noindent\textbf{Formalism used in the original work:} Agent-Based Simulation
\\
\noindent\textbf{Checklist Use:} In the SimMobility scenario, C1 (fidelity) is central, as millions of heterogeneous agents require detailed representation. C2 (human component) is minor, since policy and demand appear only in the background. C3 (dynamic interactions) is crucial, given the strong interdependence between land use, transport, and communication systems. C4 (discrete behavior) is also important, combining discrete choices with continuous flows. C5 (atomicity) has limited weight, while C6 (composite structure) is highly relevant due to structural changes such as relocation or network adaptation. C7 (heterogeneity) is essential, reflecting the diversity of agents. C8 (event sequence) plays a small role, as analysis focuses on aggregated dynamics, and C9 (feedback) is moderately relevant. Overall, the emphasis lies on structural interactions, heterogeneity, and long-term systemic impacts, with less importance given to human behavior and fine-grained event sequences.

\noindent\textbf{Result:} Agent-Based Simulation (C1, C3, C6, and C7 high).
\\
{\sloppy
\noindent\textbf{Scenario \#4 (limbo)}: Smart manufacturing and Industry 4.0 \cite{8520101}.
\\
\noindent\textbf{Brief Description:} An expert needs to evaluate the implementation of Industry 4.0 in smart manufacturing, focusing on how CPS, DTs, and decentralized decision-making can enhance productivity, quality, and sustainability.
\\
\noindent\textbf{Checklist Use:} The evaluation begins with fidelity (C1), which is partially required, as cyber-physical systems and DTs demand detailed modeling, although not all subsystems need such a fidelity. The human component (C2) remains present through operators and decentralized decision-making, but automation reduces its prominence. Dynamic interactions (C3) are important, since changes in machines, sensors, or flows propagate across the system, yet many of these effects could also be captured through aggregate models. Discrete behavior (C4) is relevant in task allocation and machine states, though continuous variables such as energy and temperature prevent discretization from being dominant. Atomicity (C5) applies to equipment treated as indivisible, while digital twins introduce internal decomposition, reinforcing the importance of composite structures (C6), as factories evolve through interdependent components that may or may not undergo structural change. Heterogeneity (C7) is evident in the variety of robots, sensors, and management platforms, though standards can lessen its impact. Event sequences (C8) emerge in production breakdowns and scheduling, but continuous flows remain equally decisive. Finally, feedback (C9) is present in the links between demand, efficiency, and resources, although it can be represented in any paradigm. Taken together, these ambiguities distribute weight across all criteria, preventing alignment with a single formalism and placing the scenario in a methodological limbo. This would require to apply the supplementary questions, trying to disambiguate and support for one of the formalisms.
\\

\noindent\textbf{5.2. Expert Evaluation.} The expert evaluation stage is still ongoing. It consisted of a questionnaire \cite{Wagner2020,wohlin2012empirical} in which the respondent watches an explanatory video about the checklist and then answers the questionnaire, assessing the checklist with respect to completeness and usability. The expert is a professor at a Canadian university. He is not a coauthor of the paper and has also more than 10 years of experience in M\&S and SE. He evaluated the initiative positively, highlighting some possible aspects to be addressed in future versions of the work, such as assessing tool support, availability of documentation and books, availability of standards, size of the community, and computational efficiency.
\\
\noindent\textbf{5.3. Limitations.} The construction of the checklist may have been biased by the authors’ experiences with the formalisms and by the support of tool-related features that do not necessarily reflect something intrinsic to the chosen formalism. 
We do not neglect the certain degree of subjectivity in judging the level of compliance with each criterion in the checklist, which may change the results depending on who uses it. 
The checklist was based on the authors’ experience and limited to three specific formalisms, which are among the most widely used in software engineering \cite{Franca2013}, and this constitutes a limitation. Moreover, we assume that the formalism chosen by the original authors in each scenario used in the evaluation is the most suitable, what can be subject to questions. We hope to receive feedback on all those factors during the conference so that we can refine it and make it available to R\&P.
\\\\
\noindent\textbf{5.4. Future Plans.} Future work includes (i) expanding the form to more experts, collecting more feedback and insights to refine and consolidate the final form of it, (ii) applying the checklist to real-world scenarios under the form of case studies, and (iii) conducting an experiment in which participants are exposed to the same scenario and asked to use the checklist to assess the level of agreement achieved among them. 

\section{Final Considerations}

This paper presented a preliminary form of a checklist to assist software engineering researchers and practitioners (R\&P) in deciding among three main formalisms to address their problems that demand modeling and simulation. A preliminary evaluation was conducted, and the results can be considered promising. We hope that this contribution can support R\&P to choose between simulation formalisms, saving their time, avoiding them to run over several studies and tutorials to realize the best formalism, while still relying on the experience of the people involved in the conception and also the experience of experts involved in their more robust forthcoming evaluation.

\newlength{\bibitemsep}\setlength{\bibitemsep}{.2\baselineskip plus .05\baselineskip minus .05\baselineskip}
\newlength{\bibparskip}\setlength{\bibparskip}{0pt}
\let\oldthebibliography\thebibliography
\renewcommand\thebibliography[1]{%
  \oldthebibliography{#1}%
  \setlength{\parskip}{\bibitemsep}%
  \setlength{\itemsep}{\bibparskip}%
}

\let\oldbibliography\thebibliography
\renewcommand{\thebibliography}[1]{\oldbibliography{#1}
\setlength{\itemsep}{0pt}}
\small

\bibliographystyle{ACM-Reference-Format}
\bibliography{sbc-template}


\begin{thebibliography}{24}


\ifx \showCODEN    \undefined \def \showCODEN     #1{\unskip}     \fi
\ifx \showISBNx    \undefined \def \showISBNx     #1{\unskip}     \fi
\ifx \showISBNxiii \undefined \def \showISBNxiii  #1{\unskip}     \fi
\ifx \showISSN     \undefined \def \showISSN      #1{\unskip}     \fi
\ifx \showLCCN     \undefined \def \showLCCN      #1{\unskip}     \fi
\ifx \shownote     \undefined \def \shownote      #1{#1}          \fi
\ifx \showarticletitle \undefined \def \showarticletitle #1{#1}   \fi
\ifx \showURL      \undefined \def \showURL       {\relax}        \fi
\providecommand\bibfield[2]{#2}
\providecommand\bibinfo[2]{#2}
\providecommand\natexlab[1]{#1}
\providecommand\showeprint[2][]{arXiv:#2}

\bibitem[Adnan et~al\mbox{.}(2016)]%
        {adnan2016simmobility}
\bibfield{author}{\bibinfo{person}{Muhammad Adnan}, \bibinfo{person}{Francisco~C Pereira}, \bibinfo{person}{Carlos Miguel~Lima Azevedo}, \bibinfo{person}{Kakali Basak}, \bibinfo{person}{Milan Lovric}, \bibinfo{person}{Sebasti{\'a}n Raveau}, \bibinfo{person}{Yi Zhu}, \bibinfo{person}{Joseph Ferreira}, \bibinfo{person}{Christopher Zegras}, {and} \bibinfo{person}{Moshe Ben-Akiva}.} \bibinfo{year}{2016}\natexlab{}.
\newblock \showarticletitle{Simmobility: A multi-scale integrated agent-based simulation platform}. In \bibinfo{booktitle}{\emph{95th Annual Meeting of the Transportation Research Board Forthcoming in Transportation Research Record}}, Vol.~\bibinfo{volume}{2}. The National Academies of Sciences, Engineering, and Medicine Washington, DC, \bibinfo{publisher}{SAGE Publishing}, \bibinfo{address}{Washington, USA}, \bibinfo{pages}{1--10}.
\newblock


\bibitem[Agalianos et~al\mbox{.}(2020)]%
        {agalianos2020discrete}
\bibfield{author}{\bibinfo{person}{K Agalianos}, \bibinfo{person}{ST Ponis}, \bibinfo{person}{E Aretoulaki}, \bibinfo{person}{G Plakas}, {and} \bibinfo{person}{O Efthymiou}.} \bibinfo{year}{2020}\natexlab{}.
\newblock \showarticletitle{Discrete event simulation and digital twins: review and challenges for logistics}.
\newblock \bibinfo{journal}{\emph{Procedia Manufacturing}}  \bibinfo{volume}{51} (\bibinfo{year}{2020}), \bibinfo{pages}{1636--1641}.
\newblock


\bibitem[Angerhofer and Angelides(2000)]%
        {899737}
\bibfield{author}{\bibinfo{person}{B.J. Angerhofer} {and} \bibinfo{person}{M.C. Angelides}.} \bibinfo{year}{2000}\natexlab{}.
\newblock \showarticletitle{System dynamics modelling in supply chain management: research review}. In \bibinfo{booktitle}{\emph{2000 Winter Simulation Conference Proceedings (Cat. No.00CH37165)}}, Vol.~\bibinfo{volume}{1}. \bibinfo{publisher}{IEEE}, \bibinfo{address}{Orlando, Florida}, \bibinfo{pages}{342--351 vol.1}.
\newblock
\href{https://doi.org/10.1109/WSC.2000.899737}{doi:\nolinkurl{10.1109/WSC.2000.899737}}


\bibitem[Caragliu and Del~Bo(2019)]%
        {caragliu2019smart}
\bibfield{author}{\bibinfo{person}{Andrea Caragliu} {and} \bibinfo{person}{Chiara~F Del~Bo}.} \bibinfo{year}{2019}\natexlab{}.
\newblock \showarticletitle{Smart innovative cities: The impact of Smart City policies on urban innovation}.
\newblock \bibinfo{journal}{\emph{Technological Forecasting and Social Change}}  \bibinfo{volume}{142} (\bibinfo{year}{2019}), \bibinfo{pages}{373--383}.
\newblock


\bibitem[de~Almeida~Molina et~al\mbox{.}(2023)]%
        {molina:2023}
\bibfield{author}{\bibinfo{person}{Sidny de Almeida~Molina}, \bibinfo{person}{Murilo Gustavo~Nabarrete Costa}, \bibinfo{person}{Abraao~Gualberto Naz{\'a}rio}, \bibinfo{person}{D{\'e}bora Maria~Barroso Paiva}, {and} \bibinfo{person}{Maria~Istela Cagnin}.} \bibinfo{year}{2023}\natexlab{}.
\newblock \showarticletitle{Cen{\'a}rios Abstratos de Tratamento de Exce{\c{c}}oes na Interoperabilidade de Processos-de-Processos de Neg{\'o}cios}. In \bibinfo{booktitle}{\emph{V MSSiS}}. \bibinfo{publisher}{SBC}, \bibinfo{address}{Curitiba, Brazil}, \bibinfo{pages}{11--20}.
\newblock


\bibitem[de~Fran{\c{c}}a and Ali(2020)]%
        {de2020role}
\bibfield{author}{\bibinfo{person}{Breno Bernard~Nicolau de Fran{\c{c}}a} {and} \bibinfo{person}{Nauman~Bin Ali}.} \bibinfo{year}{2020}\natexlab{}.
\newblock \showarticletitle{The role of simulation-based studies in software engineering research}.
\newblock In \bibinfo{booktitle}{\emph{Contemporary empirical methods in software engineering}}. \bibinfo{publisher}{Springer}, \bibinfo{address}{London, UK}, \bibinfo{pages}{263--287}.
\newblock


\bibitem[de~Fran{\c{c}}a and {Graciano Neto}(2021)]%
        {de2021opportunities}
\bibfield{author}{\bibinfo{person}{Breno Bernard~Nicolau de Fran{\c{c}}a} {and} \bibinfo{person}{Valdemar~Vicente {Graciano Neto}}.} \bibinfo{year}{2021}\natexlab{}.
\newblock \showarticletitle{Opportunities for simulation in software engineering}. In \bibinfo{booktitle}{\emph{III MSSiS}}. \bibinfo{publisher}{SBC}, \bibinfo{address}{Salvador, Brazil}, \bibinfo{pages}{50--54}.
\newblock


\bibitem[de~Fran{\c{c}}a and Travassos(2013)]%
        {Franca2013}
\bibfield{author}{\bibinfo{person}{Breno Bernard~Nicolau de Fran{\c{c}}a} {and} \bibinfo{person}{Guilherme~H. Travassos}.} \bibinfo{year}{2013}\natexlab{}.
\newblock \showarticletitle{Are We Prepared for Simulation Based Studies in Software Engineering Yet?}
\newblock \bibinfo{journal}{\emph{{CLEI} Electron. J.}} \bibinfo{volume}{16}, \bibinfo{number}{1} (\bibinfo{year}{2013}), \bibinfo{pages}{1--36}.
\newblock
\href{https://doi.org/10.19153/CLEIEJ.16.1.8}{doi:\nolinkurl{10.19153/CLEIEJ.16.1.8}}


\bibitem[{de Paula Ferreira} et~al\mbox{.}(2020)]%
        {DEPAULAFERREIRA2020}
\bibfield{author}{\bibinfo{person}{William {de Paula Ferreira}}, \bibinfo{person}{Fabiano Armellini}, {and} \bibinfo{person}{Luis~Antonio {De Santa-Eulalia}}.} \bibinfo{year}{2020}\natexlab{}.
\newblock \showarticletitle{Simulation in industry 4.0: A state-of-the-art review}.
\newblock \bibinfo{journal}{\emph{Computers \& Industrial Engineering}}  \bibinfo{volume}{149} (\bibinfo{year}{2020}), \bibinfo{pages}{106868}.
\newblock
\showISSN{0360-8352}


\bibitem[dos Santos et~al\mbox{.}(2024)]%
        {santos2024}
\bibfield{author}{\bibinfo{person}{Carlos~Henrique dos Santos}, \bibinfo{person}{José~Arnaldo Barra~Montevechi}, \bibinfo{person}{Afonso~Teberga Campos}, \bibinfo{person}{Rafael de Carvalho~Miranda}, \bibinfo{person}{José~Antonio de Queiroz}, {and} \bibinfo{person}{João~Victor Soares~do Amaral}.} \bibinfo{year}{2024}\natexlab{}.
\newblock \showarticletitle{Simulation-Based Digital Twins: An Accreditation Method}. In \bibinfo{booktitle}{\emph{2024 Winter Simulation Conference (WSC)}}. \bibinfo{publisher}{IEEE}, \bibinfo{address}{Orlando, Florida}, \bibinfo{pages}{2856--2867}.
\newblock
\href{https://doi.org/10.1109/WSC63780.2024.10838895}{doi:\nolinkurl{10.1109/WSC63780.2024.10838895}}


\bibitem[Forrester(1961)]%
        {forrester1961industrial}
\bibfield{author}{\bibinfo{person}{J.W. Forrester}.} \bibinfo{year}{1961}\natexlab{}.
\newblock \bibinfo{booktitle}{\emph{Industrial Dynamics}}.
\newblock \bibinfo{publisher}{M.I.T. Press}, \bibinfo{address}{Massachussets, USA}.
\newblock
\showISBNx{9780262060035}
\showLCCN{61017871}
\urldef\tempurl%
\url{https://books.google.com.br/books?id=4CgzAAAAMAAJ}
\showURL{%
\tempurl}


\bibitem[Jadrić et~al\mbox{.}(2020)]%
        {9245311}
\bibfield{author}{\bibinfo{person}{Mario Jadrić}, \bibinfo{person}{Maja Ćukušić}, {and} \bibinfo{person}{Dino Pavlić}.} \bibinfo{year}{2020}\natexlab{}.
\newblock \showarticletitle{Review of discrete simulation modelling use in the context of smart cities}. In \bibinfo{booktitle}{\emph{2020 43rd International Convention on Information, Communication and Electronic Technology (MIPRO)}}. \bibinfo{publisher}{IEEE}, \bibinfo{address}{Opatija, Croatia}, \bibinfo{pages}{1807--1812}.
\newblock
\href{https://doi.org/10.23919/MIPRO48935.2020.9245311}{doi:\nolinkurl{10.23919/MIPRO48935.2020.9245311}}


\bibitem[Khakifirooz et~al\mbox{.}(2018)]%
        {8520101}
\bibfield{author}{\bibinfo{person}{Marzieh Khakifirooz}, \bibinfo{person}{Dimitri Cayard}, \bibinfo{person}{Chen~Fu Chien}, {and} \bibinfo{person}{Mahdi Fathi}.} \bibinfo{year}{2018}\natexlab{}.
\newblock \showarticletitle{A System Dynamic Model for Implementation of Industry 4.0}. In \bibinfo{booktitle}{\emph{2018 International Conference on System Science and Engineering (ICSSE)}}. \bibinfo{publisher}{IEEE}, \bibinfo{address}{New Taipei City, Taiwan}, \bibinfo{pages}{1--6}.
\newblock
\href{https://doi.org/10.1109/ICSSE.2018.8520101}{doi:\nolinkurl{10.1109/ICSSE.2018.8520101}}


\bibitem[Kirkwood(1994)]%
        {Kirkwood:1994:CPM}
\bibfield{author}{\bibinfo{person}{James~R. Kirkwood}.} \bibinfo{year}{1994}\natexlab{}.
\newblock \bibinfo{booktitle}{\emph{Calculus projects for {Mathematica}}}.
\newblock \bibinfo{publisher}{McGraw-Hill Education}, \bibinfo{address}{New York, United States}. vi + 192 pages.
\newblock
\showISBNx{0-697-16736-4}
\showISBNxiii{978-0-697-16736-1}


\bibitem[Lebtag et~al\mbox{.}(2022)]%
        {Lebtag2022}
\bibfield{author}{\bibinfo{person}{Bruno Gabriel~Ara{\'{u}}jo Lebtag}, \bibinfo{person}{Paulo~Gabriel Teixeira}, \bibinfo{person}{Rodrigo~Pereira dos Santos}, \bibinfo{person}{Davi Viana}, {and} \bibinfo{person}{Valdemar Vicente~Graciano Neto}.} \bibinfo{year}{2022}\natexlab{}.
\newblock \showarticletitle{Strategies to Evolve ExM Notations Extracted from a Survey with Software Engineering Professionals Perspective}.
\newblock \bibinfo{journal}{\emph{J. Softw. Eng. Res. Dev.}}  \bibinfo{volume}{10} (\bibinfo{year}{2022}), \bibinfo{pages}{2:1--2:24}.
\newblock
\href{https://doi.org/10.5753/JSERD.2021.1939}{doi:\nolinkurl{10.5753/JSERD.2021.1939}}


\bibitem[Liu and David(2024)]%
        {liu2024ai}
\bibfield{author}{\bibinfo{person}{Xiaoran Liu} {and} \bibinfo{person}{Istvan David}.} \bibinfo{year}{2024}\natexlab{}.
\newblock \showarticletitle{AI Simulation by Digital Twins: Systematic Survey of the State of the Art and a Reference Framework}. In \bibinfo{booktitle}{\emph{Proceedings of the ACM/IEEE 27th International Conference on Model Driven Engineering Languages and Systems}}. \bibinfo{publisher}{ACM}, \bibinfo{address}{Linz, Austria}, \bibinfo{pages}{401--412}.
\newblock


\bibitem[Petersen et~al\mbox{.}(2021)]%
        {PETERSEN2021103541}
\bibfield{author}{\bibinfo{person}{Kai Petersen}, \bibinfo{person}{Jan Carlson}, \bibinfo{person}{Efi Papatheocharous}, {and} \bibinfo{person}{Krzysztof Wnuk}.} \bibinfo{year}{2021}\natexlab{}.
\newblock \showarticletitle{Context checklist for industrial software engineering research and practice}.
\newblock \bibinfo{journal}{\emph{Computer Standards \& Interfaces}}  \bibinfo{volume}{78} (\bibinfo{year}{2021}), \bibinfo{pages}{103541}.
\newblock
\showISSN{0920-5489}
\href{https://doi.org/10.1016/j.csi.2021.103541}{doi:\nolinkurl{10.1016/j.csi.2021.103541}}


\bibitem[Schlager(2008)]%
        {schlager2008hardware}
\bibfield{author}{\bibinfo{person}{Martin Schlager}.} \bibinfo{year}{2008}\natexlab{}.
\newblock \bibinfo{booktitle}{\emph{Hardware-in-the-loop simulation}}.
\newblock \bibinfo{publisher}{VDM Verlag}, \bibinfo{address}{Saarbrücken, Germany}.
\newblock


\bibitem[Singh and Singh(2009)]%
        {singh2009system}
\bibfield{author}{\bibinfo{person}{V.P. Singh} {and} \bibinfo{person}{V.P. Singh}.} \bibinfo{year}{2009}\natexlab{}.
\newblock \bibinfo{booktitle}{\emph{System Modeling and Simulation}}.
\newblock \bibinfo{publisher}{New Age International (P) Limited}, \bibinfo{address}{Delhi, India}.
\newblock
\showISBNx{9788122423860}
\urldef\tempurl%
\url{https://books.google.com.br/books?id=7LvkDg1yBgsC}
\showURL{%
\tempurl}


\bibitem[Wagner et~al\mbox{.}(2020)]%
        {Wagner2020}
\bibfield{author}{\bibinfo{person}{Stefan Wagner}, \bibinfo{person}{Daniel M{\'{e}}ndez}, \bibinfo{person}{Michael Felderer}, \bibinfo{person}{Daniel Graziotin}, {and} \bibinfo{person}{Marcos Kalinowski}.} \bibinfo{year}{2020}\natexlab{}.
\newblock \showarticletitle{Challenges in Survey Research}.
\newblock In \bibinfo{booktitle}{\emph{Contemporary Empirical Methods in Software Engineering}}, \bibfield{editor}{\bibinfo{person}{Michael Felderer} {and} \bibinfo{person}{Guilherme~Horta Travassos}} (Eds.). \bibinfo{publisher}{Springer}, \bibinfo{address}{Heidelberg, Germany}, \bibinfo{pages}{93--125}.
\newblock
\href{https://doi.org/10.1007/978-3-030-32489-6\_4}{doi:\nolinkurl{10.1007/978-3-030-32489-6\_4}}


\bibitem[Werlang(2013)]%
        {werlang2013simulaccao}
\bibfield{author}{\bibinfo{person}{P Werlang}.} \bibinfo{year}{2013}\natexlab{}.
\newblock \emph{\bibinfo{title}{Simula{\c{c}}{\~a}o da curva de crescimento do mycobacterium tuberculosis utilizando sistemas multiagentes}}.
\newblock \bibinfo{thesistype}{Ph.\,D. Dissertation}. \bibinfo{school}{Master thesis, Programa de P{\'o}s-Gradua{\c{c}}{\~a}o em Modelagem Computacional, FURG~}.
\newblock


\bibitem[Wohlin et~al\mbox{.}(2012)]%
        {wohlin2012empirical}
\bibfield{author}{\bibinfo{person}{Claes Wohlin}, \bibinfo{person}{Per Runeson}, \bibinfo{person}{Martin H\"{o}st}, \bibinfo{person}{Magnus~C. Ohlsson}, \bibinfo{person}{Bj\"{o}rn Regnell}, {and} \bibinfo{person}{Anders Wessl{\'e}n}.} \bibinfo{year}{2012}\natexlab{}.
\newblock \showarticletitle{Empirical Strategies}.
\newblock In \bibinfo{booktitle}{\emph{Experimentation in Software Engineering}}. \bibinfo{series}{International Series in Software Engineering}, Vol.~\bibinfo{volume}{6}. \bibinfo{publisher}{Springer Science \& Business Media}, \bibinfo{address}{Heidelberg, Germany}, \bibinfo{pages}{9--36}.
\newblock
\showISBNx{978-3642290435}


\bibitem[Zeigler et~al\mbox{.}(2016)]%
        {Zeigler2016BookGuideMSSoS}
\bibfield{author}{\bibinfo{person}{Bernard Zeigler}, \bibinfo{person}{Hessam~S. Sarjoughian}, \bibinfo{person}{Raphal Duboz}, {and} \bibinfo{person}{Jean-Christophe Soulie}.} \bibinfo{year}{2016}\natexlab{}.
\newblock \bibinfo{booktitle}{\emph{Guide to Modeling and Simulation of Systems of Systems} (\bibinfo{edition}{1st} ed.)}.
\newblock \bibinfo{publisher}{Springer Publishing Company, Incorporated}, \bibinfo{address}{Heidelberg, Germany}.
\newblock
\showISBNx{1447169336}


\bibitem[Zeigler(1972)]%
        {Zeigler72}
\bibfield{author}{\bibinfo{person}{Bernard~P. Zeigler}.} \bibinfo{year}{1972}\natexlab{}.
\newblock \showarticletitle{Toward a Formal Theory of Modeling and Simulation: Structure Preserving Morphisms}.
\newblock \bibinfo{journal}{\emph{J. {ACM}}} \bibinfo{volume}{19}, \bibinfo{number}{4} (\bibinfo{year}{1972}), \bibinfo{pages}{742--764}.
\newblock
\href{https://doi.org/10.1145/321724.321737}{doi:\nolinkurl{10.1145/321724.321737}}


\end{thebibliography}
\end{document}